\documentclass[aps,prb,twocolumn,groupedaddress]{revtex4-2}

\usepackage{amsmath}
\usepackage{graphicx}
\usepackage{dsfont}
\usepackage{xcolor}
\usepackage{amsfonts}
\usepackage{ulem}
\usepackage{enumitem}  
\usepackage{hyperref}
\usepackage{physics}
\usepackage{bbm}
\newcommand{\Bra}[1]{\left\langle #1 \right|}
\newcommand{\Ket}[1]{\left| #1\right\rangle}
\newcommand{\avg}[1]{\langle #1\rangle}

\definecolor{colorlibk}{rgb}{0.05, 0.4, 0.3} 

\begin{document}

\title{Designing arbitrary single-axis rotations robust against perpendicular time-dependent noise}

\author{Bikun Li} 
\affiliation{Department of Physics, Virginia Tech, Blacksburg, Virginia 24061, USA}
\author{F. A. Calderon-Vargas} 
\affiliation{Department of Physics, Virginia Tech, Blacksburg, Virginia 24061, USA}
\author{Junkai Zeng} 
\affiliation{Department of Physics, Virginia Tech, Blacksburg, Virginia 24061, USA}
\author{Edwin Barnes}
\email{efbarnes@vt.edu}
\affiliation{Department of Physics, Virginia Tech, Blacksburg, Virginia 24061, USA}

\begin{abstract}
Low-frequency time-dependent noise is one of the main obstacles on the road towards a fully scalable quantum computer. The majority of solid-state qubit platforms, from superconducting circuits to spins in semiconductors, are greatly affected by $1/f$ noise. Among the different control techniques used to counteract noise effects on the system, dynamical decoupling sequences are one of the most effective. However, most dynamical decoupling sequences require unbounded and instantaneous pulses, which are unphysical and can only implement identity operations. Among methods that do restrict to bounded control fields, there remains a need for protocols that implement arbitrary gates with lab-ready control fields. In this work, we introduce a protocol to design bounded and continuous control fields that implement arbitrary single-axis rotations while shielding the system from low-frequency time-dependent noise perpendicular to the control axis. We show the versatility of our method by presenting a set of non-negative-only control pulses that are immediately applicable to quantum systems with constrained control, such as singlet-triplet spin qubits.
Finally, we demonstrate the robustness of our control pulses against classical $1/f$ noise and noise modeled with a random quantum bath, showing that our pulses can even outperform ideal dynamical decoupling sequences.

\end{abstract}
	
	\maketitle
	
	\section{Introduction}
	The rapid technological advancements of the last decade in the design, fabrication, and control of solid-state qubit systems has produced two front-runners, spin qubits~\cite{Hanson2007,Zwanenburg2013,gonzalez2020scaling} and superconducting qubits~\cite{devoret2005implementing,clarke2008superconducting,Kjaergaard2019,devoret2013superconducting,wendin2017quantum}, in the race to build the first fully functional quantum computer.
	One of the features any viable approach to quantum information processing must have is the ability to implement high-fidelity quantum gates, a cornerstone for fault-tolerant quantum computation.
	Recent experiments with spin qubits, notably those encoded in the spin state of one or more electrons trapped in one or more quantum dots, have demonstrated high-fidelity single-~\cite{Kim2015b,Kawakami2016,Yoneda2017,takeda2020resonantly,cerfontaine2020closed} and two-qubit gates~\cite{brunner2011two,Veldhorst2015a,Nichol2016,Zajac2018,Watson2018,Xue2018a,Huang2018c,sigillito2019coherent}, even at temperatures above one Kelvin~\cite{Yang2019,Petit2020,Petit2020a}. Similarly, owing to their strong nonlinearity, design flexibility, and small dissipation, high-fidelity single-~\cite{gustavsson2013improving,chen2016measuring,sheldon2016characterizing,rol2017restless} and two-qubit gates~\cite{chow2012universal,barends2014superconducting,Sheldon2016,reagor2018demonstration} have been implemented with superconducting qubits. Incidentally, superconducting qubits are at the core of the online-accessible IBM Quantum Experience quantum processors, which were used, among other things, to demonstrate the experimental implementation of small error detection codes~\cite{takita2017experimental} and to perform basic quantum simulations~\cite{o2016scalable,kandala2017hardware}.

	Despite the major advancements in quantum gate implementation, decoherence due to the interaction with a noisy environment is still a main hurdle toward a fully functional quantum computer. For spin qubits, there are two types of noise sources: charge and magnetic. Charge noise stems from the coupling to nearby two-level fluctuators~\cite{Dial2013} and stray electric fields~\cite{Dijk2018}. This type of noise causes fluctuations in the tunneling and detuning energies~\cite{Paladino2014,Huang2018a}, affecting the exchange coupling and single-qubit resonance frequencies~\cite{Yoneda2017,Chan2018}. Magnetic noise is caused by fluctuations in the surrounding nuclear spin bath, which perturbs the qubit system via the hyperfine interaction. The variance of these fluctuations, however, can be reduced using a feedback scheme based on dynamic nuclear polarization~\cite{Bluhm2010a}. For superconducting qubits, charge noise and magnetic flux noise are the main sources of decoherence~\cite{oliver2013materials}. On one hand, charge noise, due to nearby charge fluctuators or charge traps in the substrate~\cite{wang2015surface,dial2016bulk}, is mainly responsible for energy relaxation but can also affect the qubit energy splitting if the ratio between Josephson and charging energies is not large enough. On the other hand, magnetic flux noise, caused by random flipping of magnetic moments of electrons on the surface of the superconductor~\cite{koch2007model,slichter2012measurement,yan2013rotating}, produces stochastic detuning of flux-tunable qubits. Noise can be reduced to a certain extent through materials and fabrication improvements, e.g., for superconducting qubits substrate cleaning reduces the number of defects~\cite{Quintana_2014}, and for silicon-based spin qubits, noise from the surrounding nuclear spin bath can be practically suppressed by isotopically purifying the substrate~\cite{Itoh2014}. The noise can be further reduced in both superconducting and spin qubits via quantum control techniques as evidenced by several experimental results~\cite{Gustavsson_2012,Martins_2016,Sheldon2016,pokharel2018demonstration,Yang_2019,cerfontaine2020closed}. This is owing to the fact that charge and magnetic noise exhibit a $1/f$-type power spectrum~\cite{bialczak20071,yan2012spectroscopy,Dial2013,freeman2016comparison,Malinowski2017}, i.e., most of the noise power resides at low frequencies. Therefore, both types of noise can be approximated as quasi-static and are thus responsive to engineered error mitigation techniques such as spin echo~\cite{Hahn_Echo}, dynamical decoupling sequences~\cite{Carr_1954,Meiboom1958,Gullion1990,Viola1999,XY8_PRL.90.037901,UDD,Universality_UDD,QDD,NUDD}, composite pulses~\cite{Levitt_1986,Wimperis1994,CORPSE,Jones2010,Merrill2012,Wang2014,Calderon-Vargas2016}, optimal control theory~\cite{konnov1999global,Palao2002,Khaneja2005,Doria_2011,Caneva_2011}, pulse engineering~\cite{Khodjasteh_DECG,Khodajasteh_Viola_2010,Barnes2012a,Economou2015}, geometric-formalism-based pulse control~\cite{Barnes2015,Zeng2018b,Zeng_2018,Zeng_2019,buterakos2020geometrical}, etc. For quantum computing applications, however, cancelling quasi-static noise is not enough to implement gate operations with errors below the quantum error correction threshold; time-dependent non-Markovian noise must also be corrected.

	The effects of time-dependent noise can be mitigated to a certain extent via dynamical decoupling~\cite{CDD_2005}, dynamically corrected gates~\cite{Khodjasteh2012}, or composite pulses~\cite{Kabytayev2014}. The performance of these approaches, or any noise mitigation technique in general, can be quantitatively characterized by their filter functions~\cite{Cywinski_PRB,Biercuk_2011}, which show that time-dependent noise is mitigated as long as the correlation time scale of the noise is longer than the control time scale of these error-correcting sequences.  The aforementioned approaches, however, are based on instantaneous `bang-bang' pulses. Implementing these Dirac-delta-like pulses would require infinite power, and thus they can only be approximated by real physical pulses, with finite width and (almost-)always-on control fields, resulting in a subpar noise cancellation. Previous works have studied the use of bounded-strength controls in both dynamically decoupling sequences~\cite{Uhrig_2010} and dynamically corrected gates~\cite{Khodjasteh2009b,Green2012,Green_2013}, improving existing sequences by using realistic pulses. The use of always-on control fields was studied in Refs.~\cite{Cody_Jones_2012,Cai2012}, where they introduced schemes for dynamical decoupling sequences with bounded, continuous control fields. So far, however, there remains a need for a general method to design arbitrary gates that suppress time-dependent noise using bounded-strength, always-on control fields.
	
	In this work, we combine the filter function formalism and the geometric formalism~\cite{Zeng_2019} together with numerical techniques to design bounded continuous control fields that produce arbitrary single-axis rotations robust to time-dependent noise perpendicular to the control axis.  We show that these smooth error-correcting pulses are equivalent to sequences of closed curves in the geometric formalism. The flexibility and efficacy of our method is demonstrated for both classical $1/f$ noise and noise modeled by a quantum bath. Moreover, our method can be used in systems with restrictive control fields like the non-negative-only control available in singlet-triplet spin qubits~\cite{Petta2005,Wang2014}.

	This paper is organized as follows: Sec.~\ref{sec:hamiltonia and geometric formalism} introduces a modified version of the geometric formalism for time-dependent noise. Sec.~\ref{sec:numerical scheme} presents a scheme to design smooth continuous pulses that produce arbitrary single-axis rotations resistant to time-dependent noise. Then,  Sec.~\ref{sec:Realistic time-dependent noise} examines how well the proposed smooth pulses perform against realistic noise power spectra. We conclude in Sec.~\ref{Sec:conclusion}.

\section{Noisy Hamiltonian and geometric formalism}\label{sec:hamiltonia and geometric formalism}
	We consider a general two-level system, coupled to an external quantum bath, with initial state  $\Ket{\psi_0}\Bra{\psi_0}\otimes\rho_B$ and Hamiltonian given by
	\begin{equation}\label{eq:Hamiltonian0}
	    H(t) = \frac{\Omega(t)}{2}\sigma_x\otimes \mathbbm{1}
	    + \lambda \sigma_z \otimes B
	    + \mathbbm{1}\otimes H_B\;,
	\end{equation}
	where $\Omega(t)$ is the control field, $\sigma_\alpha\;(\alpha = x,y,z)$ are Pauli matrices acting on the two-level system, and $\lambda$ is a coupling term. The bounded operators $B$ and $H_B\propto \omega_B$  act on the environment, i.e. a generic quantum bath. The coupling to the environment ($\lambda$) is assumed sufficiently small that it induces a slow coherence decay within a determined time interval $t\in [0,T]$. Here, $T$ is the time it takes to implement a target operation $U_{target}\equiv U(T)$, where $U(t)$ is the evolution operator generated by $H(t)$. 
We want to determine what functions for $\Omega(t)$ can implement a desired operation while canceling the leading-order errors due to the coupling $\lambda$. To that end, it is convenient to work in the interaction picture, which is defined with respect to $U_c(t)=e^{-i\frac{\phi(t)}{2}\sigma_x}\otimes e^{-iH_Bt}$, where $\hbar = 1$, $U(t) = U_c(t)U_I(t)$, $\phi(t)=\int_0^t\Omega(\tau)\dd{\tau}$ is the target rotation, and $U_I(t) = \hat{\mathcal{T}}\exp[-i\int_0^t H_I(\tau)\dd{\tau}]$ is the interaction picture evolution operator.
The interaction  Hamiltonian $ H_I$ is written as:
	\begin{equation}\label{eq:Hinteraction}
	    H_I(t) =\lambda e^{i\frac{\phi(t)}{2}\sigma_x}\sigma_z e^{-i\frac{\phi(t)}{2}\sigma_x}\otimes B_I(t),
	\end{equation}
	where $B_I(t) = e^{iH_Bt}Be^{-iH_Bt}$, which sometimes is simply replaced by a classical random time-dependent noise term~\cite{Cywinski_PRB, Wang_Liu_No_go_PhysRevA.87.042319}. 	
The dynamics of $B_I(t)$ is assumed to be `slow', i.e., the Fourier transform $S(\omega)$ of the two-point noise correlation function $\avg{B_I(t)B_I(0)}$ should concentrate around $\omega = 0$, with a bandwidth $\omega_B\ll T^{-1}$.

We are interested in suppressing the leading-order effect of noise, and so we treat the interaction Hamiltonian $H_I$ as a perturbation. We  can then perform a Magnus expansion of $U_I(T)$:
	\begin{equation}\label{eq:MagnusExpansion}
	    \begin{aligned}
	    U_I(T)&=\exp\left[-i( M_1(T) + M_2(T) +\cdots)\right]\\
	    &=1- i [M_1(T)+M_2(T)]-\frac{1}{2}M_1(T)^2 + \mathcal{O}(\lambda^3),
	    \end{aligned}
	\end{equation}
	where $M_1(t)=\int_0^tH_I(\tau)\dd{\tau}$ and $M_2(t)=\frac{1}{2}\int_0^t\dd{\tau}\int_0^{\tau}\dd{\tau}'[H_I(\tau),H_I(\tau')]$. 
	In the geometric formalism~\cite{Zeng2018b,Zeng_2018,Zeng_2019}, we treat $B_I$ as quasi-static during the gate operation, which allows us to map $M_1(t)$ to a plane curve $\vec{C}(t)$ with tangent vector $(\cos\phi,\sin\phi)$ and curvature $\Omega(t)=\dot{\phi}$. In this point of view, a closed curve is equivalent to $M_1(T)$ vanishing. Moreover, $M_2(T)$ is proportional to the net area enclosed by $\vec{C}(t)$~\cite{Zeng2018b}, which can be made to vanish by properly shaping the curve. Thus, pulses that cancel noise can be obtained by drawing closed curves and reading off the curvature. When $B_I(t)$ cannot be approximated as quasi-static, however, the direct application of the geometric formalism is not possible. Nonetheless, a geometric picture can be recovered by first combining the Magnus expansion of $U_I(T)$ and the expression for average gate fidelity~\cite{BOWDREY_Fidelity} (see Appendix~\ref{app:derivation_Eq4}): 
	\begin{equation}\label{eq:fidelity}
	 \begin{aligned}
	    &\mathcal{F} = \frac{1}{2} + 
	    \frac{1}{12}
	    \!\!\sum_{\alpha = x,y,z}\!\!
	    \Tr[(\sigma_\alpha\otimes \mathbbm{1})U_I(T)(\sigma_\alpha\otimes \rho_B)U_I^{\dagger}(T)]
	    \\
	& \approx 1 - \frac{2\lambda^2}{3}\!\!\int_0^T\!\!\!\!\dd{\tau_1}\!\!\int_0^T\!\!\!\!\dd{\tau_2}
	\cos[\phi(\tau_1)-\phi(\tau_2)]\avg{B_I(\tau_1)B_I(\tau_2)},
	\end{aligned}
	\end{equation}
where we have omitted higher-order terms. Here, the quantum correlator $\avg{B_I(\tau_1)B_I(\tau_2)}\equiv \mathrm{tr}(B_I(\tau_1)B_I(\tau_2)\rho_B)$ has a short correlation time $\sim \omega_B^{-1}$. Using the Fourier transform of the two-point correlation function, $\lambda^2\avg{B_I(\tau_1)B_I(\tau_2)} = \int_{-\infty}^\infty\frac{\dd{\omega}}{2\pi} S(\omega)e^{-i\omega(\tau_1-\tau_2)}$ (where we assume that the correlator is time local), the gate infidelity ($1-\mathcal{F}$) can be approximated as
	\begin{equation}\label{eq:infidelity_leading}
	    1-\mathcal{F} \approx 
	    \frac{1}{3}\int_{-\infty}^\infty\frac{\dd{\omega}}{2\pi} S(\omega)
	    F(\omega,T),
	\end{equation}
	in which $F(\omega,T)\equiv |f(\omega,T)|^2 + |f(-\omega,T)|^2$  with
	\begin{equation}\label{eq:fdef}
	    f(\omega,T)\equiv \int_0^T e^{i[\phi(\tau)-\omega \tau]}\dd{\tau},
	\end{equation}
is the filter function.
Note	 that the filter function in \eqref{eq:infidelity_leading} encapsulates only the lowest-order nontrivial effect of the control field on the gate infidelity in a ever-changing noisy environment, whereas for Gaussian noise the filter function gives an exact representation~\cite{Cywinski_PRB}.
In order to minimize Eq.~\eqref{eq:infidelity_leading}, 	we need to make the filter function, $F(\omega,T)$ , as small as possible at frequencies where $S(\omega)$ is maximal. 
Hereafter, we consider the case where the noise power spectrum has its maximum at $\omega_0=0$, but it can be easily generalized to a non-zero $\omega_0$.
	
Minimizing the filter function implies the minimization of $f(\pm\omega,T)$, Eq.~\eqref{eq:fdef}, i.e. making $f(\pm\omega,T)$ as small as $\mathcal{O}(\omega^kT^{k+1})$ around $\omega=0$, where $k$ is a positive integer. To this end, we want the first $k$-th derivatives of $f(\pm\omega,T)$ to vanish at the origin:
	\begin{equation}\label{eq:vanishingder}
	\begin{aligned}
	0&=(i\partial_{\omega T})^\ell [T^{-1}f(\pm \omega,T)]_{\omega T=0} \\
	 &= 
	    \int_0^1 s^\ell e^{i\phi(Ts)}\dd{s} =\sum_{m=0}^\ell \frac{(-1)^m \ell !}{(\ell - m)!} r_m(1),
	    \\
	\end{aligned}
	\end{equation}
	with $0\le \ell < k$. The third line is obtained via integration by parts, where $r_m(s)\; (0\le s \le 1)$ is a complex function defined as
	\begin{equation}\label{eq:vanishingintegrals}
	    r_m(s) = \int_0^{s_0}\dd{s_1}\int_0^{s_1}\dd{s_2}\cdots\int_0^{s_{m}}\dd{s_{m+1}} e^{i\phi(Ts_{m+1})},
	\end{equation}
	where $s_0\equiv s$.
	All $r_m(s)$ are required to be zero at $s=1$ in order to fulfill Eq.~\eqref{eq:vanishingder}.
\begin{figure}[h]
    	\centering
    	\includegraphics[width=\linewidth]{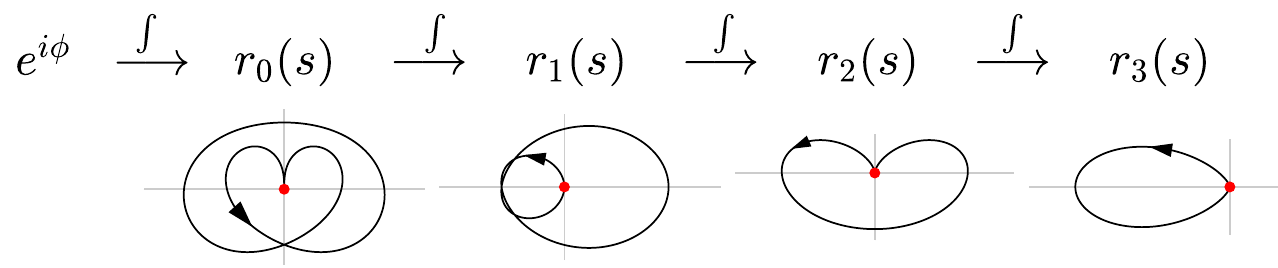}
    	\caption{An illustration of a sequence of closed curves $\{r_m\}$, where the red dots represent the starting and ending points of each curve. Each curve in the sequence is obtained by integrating the previous. $r_0(s)$ is given by the integral of $e^{i\phi(Ts)}$. In this example, the integral of $r_3(s)$ is no longer a closed curve, so the sequence terminates and thus contains four curves in total.
    	}
    	\label{fig:rmsequence}
    \end{figure}	
This expression motivates introducing a modified version of the geometric formalism	in which we treat each $r_m(s)$ in Eq.~\eqref{eq:vanishingintegrals} as a closed curve on a complex plane (see Fig.~\ref{fig:rmsequence}). The filter function $F\sim |f|^2$ is suppressed up to order $\mathcal{O}(\omega^{2k}T^{2k+2})$ about $\omega=0$ if and only if the $k$ curves in this sequence are all closed. Thus, requiring the integrals in Eq.~\eqref{eq:vanishingintegrals} to vanish can be viewed as an extension of the closed curve argument for quasi-static noise~\cite{Zeng_2018}, which simply requires that the first integral ($m=0$) vanishes. Notice that the curves in this sequence are all related through differentiation: $r_\ell'(s)=r_{\ell-1}(s)$. Therefore, the task of finding pulses that implement gates while canceling low-frequency noise to order $\mathcal{O}(\omega^{2k}T^{2k+2})$ is equivalent to finding a closed curve $r_{k-1}(s)$ such that its first $k-1$ derivatives are also themselves closed curves. In addition to being closed, the last curve obtained from this sequence of derivatives, $r_0(s)$, must also have the property that $|r_0'(s)|=1$, since $r_0(s)$ is the integral of a phase (see Eq.~\eqref{eq:vanishingintegrals}). If such a sequence of closed curves can be found, then a robust pulse can be obtained from the curvature of $r_0$. Moreover, a desired target rotation can be obtained by designing the hierarchy such that $r_0$ exhibits a cusp at the origin. The opening angle of this cusp determines the rotation angle $\theta$ about the $x$-axis~\cite{Zeng2018b}. For arbitrary $k$, it is not a simple task to find such curve sequences. While it is relatively straightforward to find closed curves $r_{k-1}(s)$ such that the first $k-1$ derivatives are also closed, it is more challenging to design $r_{k-1}(s)$ such that $|r_0'(s)|=1$ is satisfied. In the next section, we show how this problem can be circumvented by parameterizing $\phi(t)$ appropriately and using a numerical recipe to obtain control fields that suppress low-frequency noise to arbitrary order $k$.

	\section{Numerical scheme to produce robust smooth pulses}\label{sec:numerical scheme}
	
	\begin{figure*}
    	\centering
    	\includegraphics[width=0.7\linewidth]{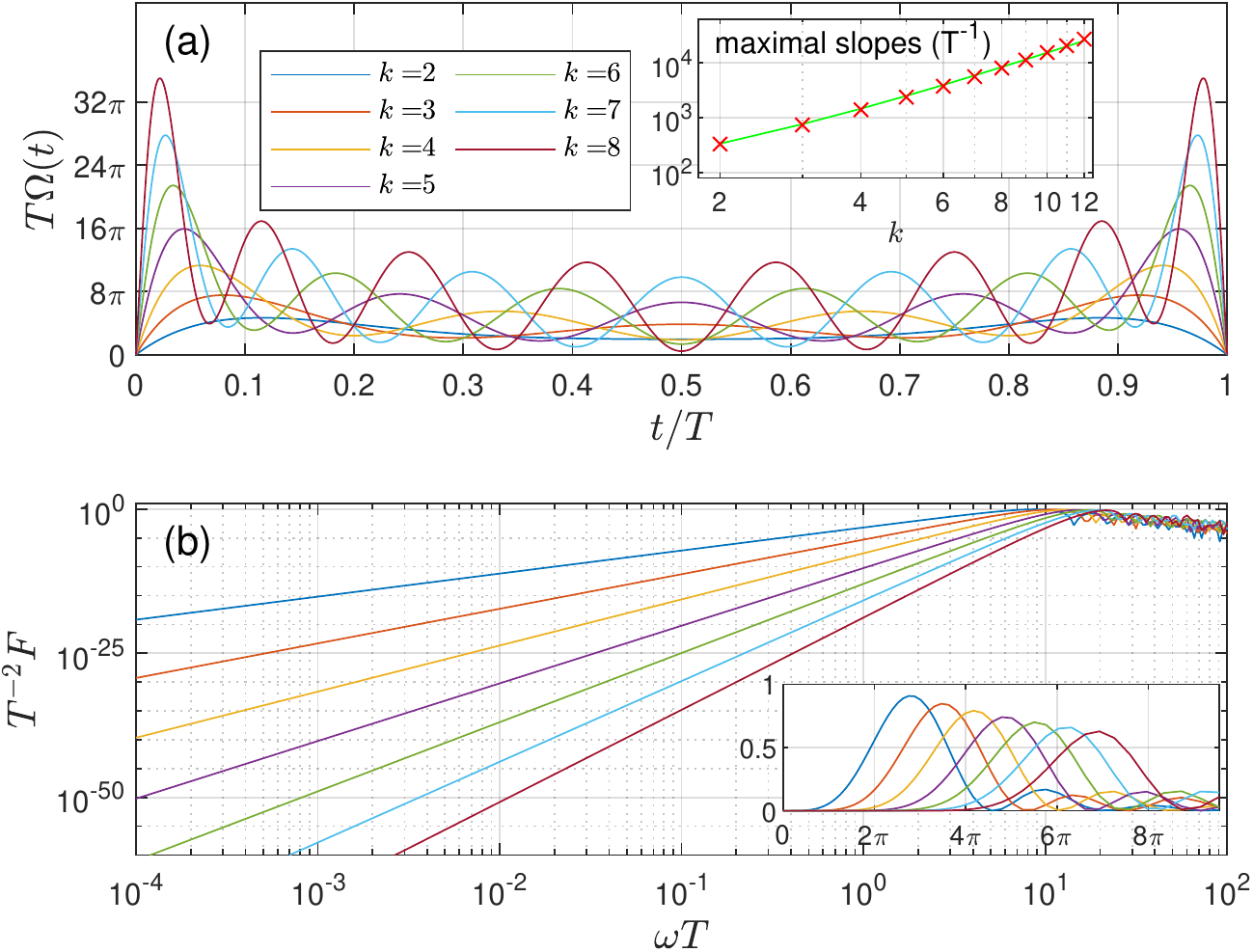}
    	\caption{(a) Seven control pulses (colored lines) that achieve $T^{-1}f(\omega,T)=\mathcal{O}(\omega^kT^k)$ noise cancellation. The $k$th pulse implements an $x$-rotation by angle $(k+1)\pi$.
    	The inset shows the maximum slopes (bandwidths) that are required for different $k$. These exhibit a $\sim k^3$ dependence at small $k$ (green line).
    	(b) The filter functions for the control pulses in (a) (colored lines). The asymptotic slopes of the filter functions are $\approx 2k$, indicating the $\mathcal{O}(\omega^{2k}T^{2k+2})$ suppression near $\omega T=0$. The inset shows the same filter functions on a linear scale.
    	}
    	\label{fig:Omegak2to8}
    \end{figure*}
	
	In this section, we develop a numerical method to obtain pulses that satisfy the closed-curve constraints in Eqs.~\eqref{eq:vanishingder} and \eqref{eq:vanishingintegrals}. We start by setting $x=2s-1$ and combining terms in Eq.~\eqref{eq:vanishingder}, which converts the error-cancellation constraints into a more convenient form:
	$\eta_\ell = \int_{-1}^{1}x^\ell e^{i\phi[T(x+1)/2]}\dd{x}=0$,
	with $0\le \ell <k$.
	Now, the key ansatz of this work is to treat the target rotation as an odd polynomial with $N$ unknown coefficients:
	\begin{equation}\label{eq:ansatz}
	    \phi\big[T(x+1)/2\big]=p_1 x + p_3 x^3 + \cdots + p_{2N-1} x^{2N-1}.
	\end{equation}
 The advantage of this ansatz is twofold: first, the control field $\Omega(t)$ is symmetric about $t=T/2$, which makes $\eta_\ell$ a real function of the $p_i$ up to a global phase, and second, there is a recursive derivative relation, $\partial \eta_\ell/\partial p_{2j-1} = i\eta_{\ell + 2j-1}$, that is useful in what follows. 
	The pulses should implement arbitrary $X_\theta$ gates in a time frame $0<t<T$ and, therefore, $\phi$ must satisfy the boundary conditions  $\phi|_{x=1}=\theta/2$ and $\partial_x\phi|_{x=1}=0$.
 In total, we therefore have $k+2$ real functions involving $p_{2j-1}$ unknowns:
	\begin{equation}\label{eq:constrFunc}
	    \begin{aligned}
	        G_\ell &= i^{\ell-1} \eta_{\ell-1},\quad \mathrm{with} \quad 1\le \ell\le k\\
	        G_{k+1} &= \bigg(\sum_{j=1}^N p_{2j-1}\bigg)-\frac{\theta}{2},\\ 
	        G_{k+2} &=\sum_{j=1}^N (2j-1)p_{2j-1},
	    \end{aligned}
	\end{equation}
	which are all required to vanish. 
	Notice that by setting $N=k+2$, the numerical solutions can be obtained efficiently through iteration (damped Newton method):
	\begin{equation}\label{eq:dampedNewton}
	p_{2j-1}^{(n+1)} = p_{2j-1}^{(n)} - \alpha\sum_{m=1}^{N}[J(p^{(n)})]^{-1}_{j,m} G_m(p^{(n)}),
    \end{equation}
	where $0<\alpha\le 1$ is the damping parameter, and $J$ is the non-singular Jacobian (square) matrix $J_{\ell,j} = \partial G_\ell/\partial p_{2j-1}$, which can be easily evaluated via the aforementioned recursive derivative relation, $\partial \eta_\ell/\partial p_{2j-1} = i\eta_{\ell + 2j-1}$.
	Now, if a proper value for $\alpha$ is chosen and the initial guess for $p^{(0)}$ is suitable, then after some number of iterations the term $\|G_m(p^{(n_f)})\|$ will inevitably fall under a predefined convergence threshold $\epsilon$, signaling that the concomitant $\widetilde{p}_{2j-1}\equiv p^{(n_f)}_{2j-1}$ is a valid solution. The control pulse is then finally given by $\Omega(\tau)=\partial_\tau \phi(\tau)|_{p=\widetilde{p}}$, which is a polynomial of degree $2k+2$.
	
	\begin{figure*}
    	\centering
    	\includegraphics[width=0.85\linewidth]{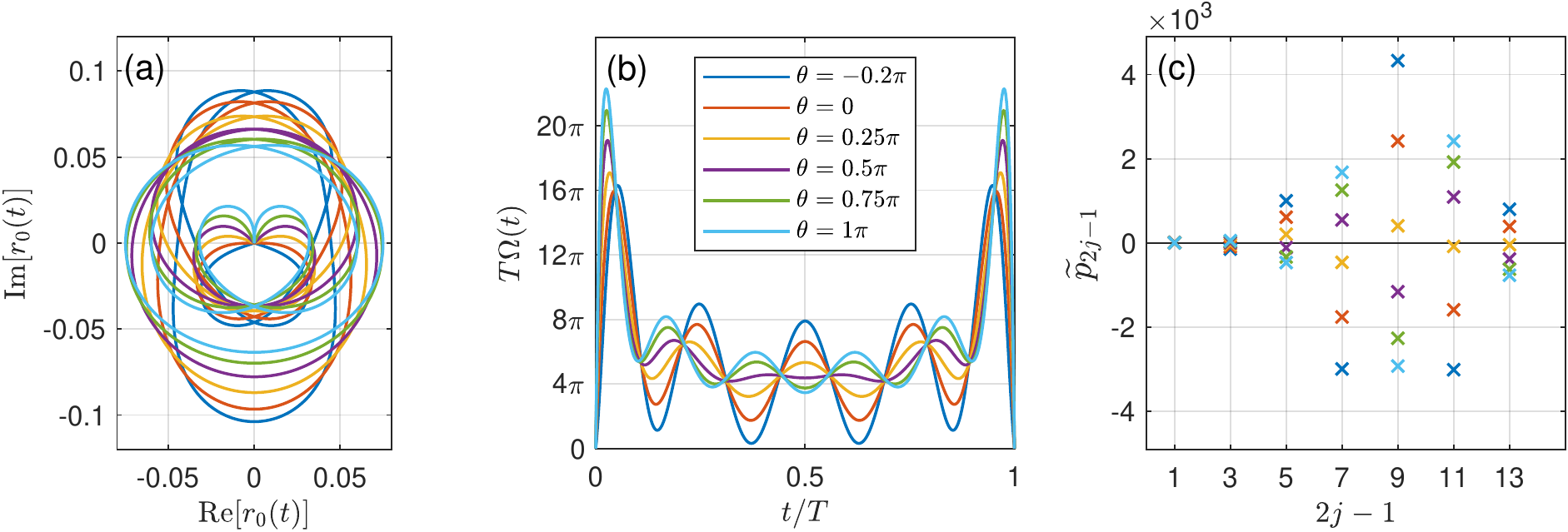}
    	\caption{(a) Closed curves $r_0(t)$ that generate a hierarchy of closed curves $r_k(t)$ up to $r_4(t)$ (see e.g., Fig.~\ref{fig:rmsequence}). The opening angles of the cusps at the origin determine the rotation angles $\theta$. (b) Pulses that implement $x$-rotations of various angles with $k=5$ time-dependent noise suppression. (c) The polynomial coefficients that define the control pulses shown in (b).
    	}
    	\label{fig:k5 var theta}
    \end{figure*}
	
	With the numerical scheme introduced above, one can easily obtain solutions for different sets of parameters. Even though we have extensively explored the solution space, here we focus on solutions that are obtained from the initial values $p_j^{(0)} = \delta_{1j}[(k+1)\pi+\theta]/2$, which correspond to a square pulse. Choosing $k\le 5$, $\alpha=1$, and threshold $\epsilon=10^{-30}$, iteration of Eq.~\eqref{eq:dampedNewton} converges to valid solutions within $100$ steps. For $k$ larger than 5, we find that the damping parameter $\alpha$ must be reduced to obtain suitable solutions. Overall, these types of solutions converge into non-negative control pulses, which are important for qubit systems with constrained control like singlet-triplet qubits~\cite{Petta2005,Wang2014}. Fig.~\hyperref[fig:Omegak2to8]{\ref*{fig:Omegak2to8}(a)} shows several non-negative control pulses obtained with $\theta = (k+1)\pi$ and $2\le k\le 8$. The inset of Fig.~\hyperref[fig:Omegak2to8]{\ref*{fig:Omegak2to8}(a)} presents, as a function of $k$, the maximal slopes of the pulses, which are related to the maximum bandwidth needed to implement them with a waveform generator. Their rate of change is proportional to $\sim k^3$.
	Fig.~\hyperref[fig:Omegak2to8]{\ref*{fig:Omegak2to8}(b)} and its inset present the filter functions, on both logarithmic and linear scales, of the control pulses presented in Fig.~\hyperref[fig:Omegak2to8]{\ref*{fig:Omegak2to8}(a)}. Note that the asymptotic slopes of the filter functions are approximately equal to $2k$, which indicates the suppression of $\mathcal{O}(\omega^{2k}T^{2k+2})$ terms near $\omega T=0$. Finally, Fig.~\hyperref[ fig:k5 var theta]{\ref*{fig:k5 var theta}} showcases the flexibility of our scheme in producing arbitrary $x$-rotations. Setting $k=5$, we repeat the above iterative procedure to obtain noise-suppressing pulses for a range of rotation angles $\theta$. In Fig.~\hyperref[ fig:k5 var theta]{\ref*{fig:k5 var theta}}(a), we show the resulting set of curves $r_0(t)$, from which the pulses in Fig.~\hyperref[ fig:k5 var theta]{\ref*{fig:k5 var theta}}(b) can be obtained from the curvatures. Each of these curves is at the bottom of a closed curve hierarchy that starts at the top with $r_4(t)$, analogous to what is shown in Fig.~\ref{fig:rmsequence} for $k=3$. Fig.~\hyperref[ fig:k5 var theta]{\ref*{fig:k5 var theta}}(b) also shows that the pulses remain non-negative across a wide range of rotation angles $\theta$. The parameters $\tilde{p}_{2j-1}$ used to construct these curves and pulses are given in Fig.~\hyperref[ fig:k5 var theta]{\ref*{fig:k5 var theta}}(c).

    We must point out that although with our method one can find solutions for a wide range of parameters, the solution space is rather complicated due to the nonlinear form of the equations in \eqref{eq:constrFunc}. Moreover, as $k$ increases, it becomes more difficult to obtain good solutions even after setting the damping parameter to $\alpha < 10^{-2}$. This is partly because the Jacobian matrix becomes almost singular for large $k$, which in turn requires an increase in precision to retain accuracy in the calculations.

	\section{Smooth continuous pulses  under realistic time-dependent noise}\label{sec:Realistic time-dependent noise}
	We have shown, so far, that smooth continuous control pulses can be designed to perform arbitrary $x$-rotations and, at the same time, shield the system from time-dependent noise. In this section we characterize the performance of the smooth control pulses against realistic noise power spectra. To this end, we calculate the gate infidelity in the presence of, first, a random quantum bath and, then, a classical $1/f^{\alpha}$ noise.
	
	\begin{figure*}
    	\centering
    	\includegraphics[width=0.6\linewidth]{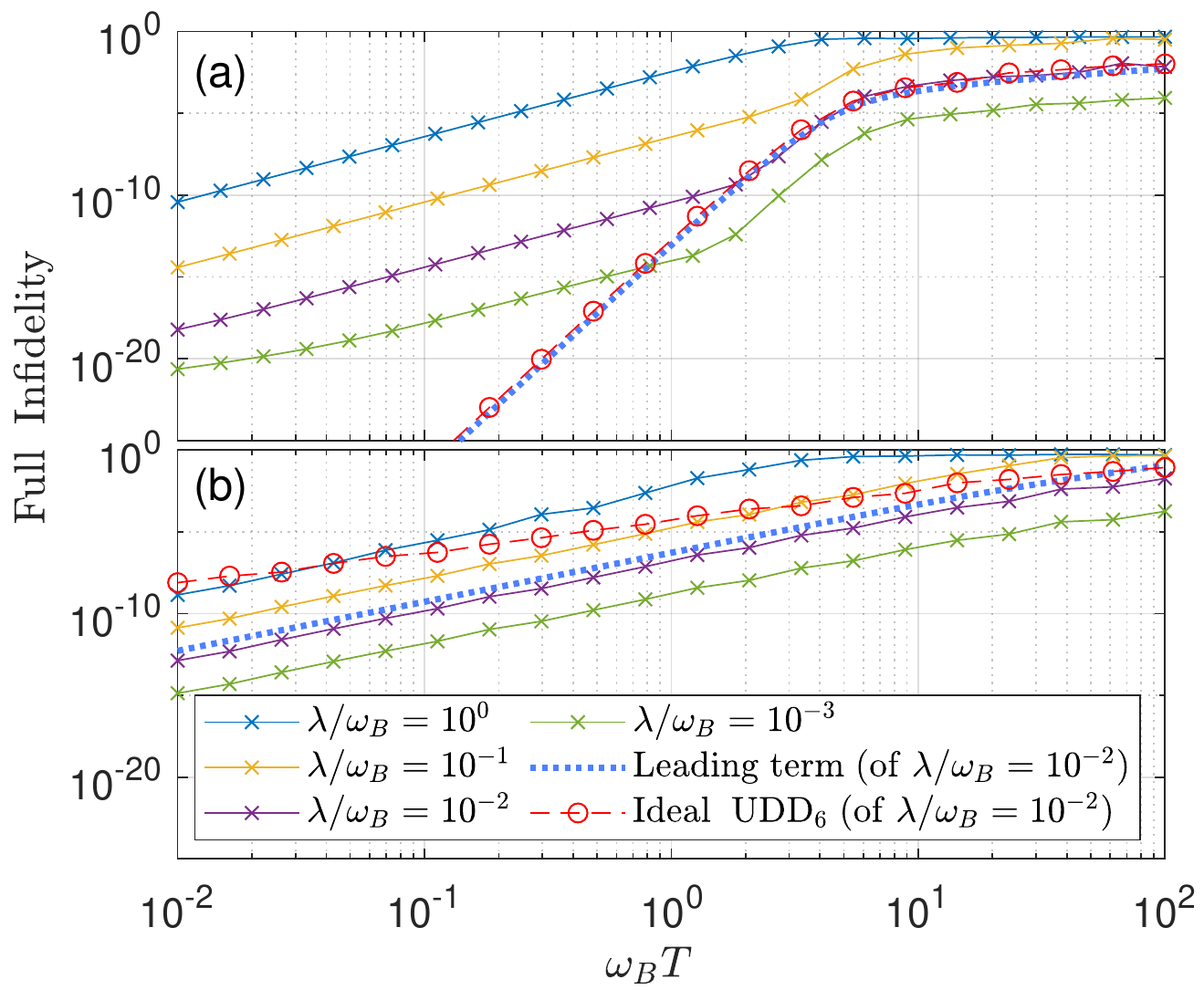}
    	\caption{The full infidelities for an $X_{\pi}$ operation using the pulse from Fig.~\ref{fig:Omegak2to8} with $k=6$ for (a) VRQB bath, and (b) classical noise $\eta(t)$ with a soft high-frequency cutoff. Each curve is plotted with varying control gate time $T$, but fixed $\omega_B$ and $\lambda/\omega_B$. The dash curve shows the leading-order contribution of the infidelity for $\lambda/\omega_B = 10^{-2}$, which has an asymptotic slope $2k+2=14$.  As an identity operation, the Uhrig dynamical decoupling sequence ($\mathrm{UDD}_6$)~\cite{UDD} is also included for comparison purposes.
    	}
    	\label{fig:full infid plot}
    \end{figure*}
	
	\paragraph{Virtual Random Quantum Bath (VRQB)}
	Since the actual environment surrounding the qubit system is usually a complicated many body quantum system, we use a bounded operator ($m$-levels) to simulate the quantum bath. 
	Therefore, we take two independent $m\times m$  random matrices $W_{1},W_{2}$ from a Gaussian unitary ensemble (GUE) and use them to define the bounded operators $B = W_1/\sqrt{m}$ and $H_B = \omega_B W_2/\sqrt{m}$, where the bath density operator is $\rho_B \propto e^{-\beta H_B}$. Regardless of the complicated dynamics of $B_I(t)$, the statistical property of the noise power spectrum has an analytic form in the thermodynamic limit $m\to \infty$ (see Appendix~\ref{app:derivation_Eq12}):
	\begin{equation}\label{eq:S_omg_VQRB}
	    S(\omega) = \frac{2\pi\lambda^2}{Z\omega_B}\int_{-\infty}^\infty  p(u)p(u -\omega/\omega_B)e^{-\beta\omega_B u}\dd{u},
	\end{equation}
	where $p(u)=(2\pi  )^{-1}\sqrt{4-u^2}$ with $|u|< 2$, and $p(u)=0$ with $|u|\ge 2$ are given by \textit{Wigner's semicircle law}. $Z\equiv \int_{-2}^2\dd{\lambda} p(\lambda) e^{-\beta\omega_B \lambda}$ is a normalization factor. Due to the piecewise $p(\omega)$ distribution, there are hard high-frequency cutoffs at $\omega = \pm 4\omega_B$.

	In Fig.~\hyperref[fig:full infid plot]{\ref*{fig:full infid plot}(a)}, the results with infinite temperature ($\beta = 0$) are computed for the $k=6$ pulse presented in Fig.~\ref{fig:Omegak2to8}, where full infidelities are presented together with the leading infidelity term, Eq.~\eqref{eq:infidelity_leading}. 
	According to dimensional analysis, each term in the expansion of the infidelity has the form: $\propto\lambda^\nu T^{\mu} \omega_B^{\mu-\nu} (\nu \ge 2)$; our leading term, Eq.~\eqref{eq:infidelity_leading}, only involves terms with $\nu=2$, and $\mu \ge 2k+2$ due to our optimized filter function $\mathcal{O}(\omega^{2k}T^{2k+2})$.
	As a result, these $\nu=2$ terms dominate when $\lambda/\omega_B\ll 1$ and $T\gg \omega_B^{-1}$.
	One can clearly see in Fig.~\hyperref[fig:full infid plot]{\ref*{fig:full infid plot}(a)} that in the region $T > \omega_B^{-1}$, the full infidelity closely follows Eq.~\eqref{eq:infidelity_leading}.

    \paragraph{Classical $1/f^\alpha$ noise}
    In almost all solid-state qubit platforms, fluctuations of various system parameters characterized by a $1/f$ power spectral density have been observed.
    Here, we consider a $1/f^\alpha$ power spectral density that remains nonzero in the high-frequency domain.
    To efficiently simulate the soft high-frequency cutoff of the noise power spectrum, we replace $B_I(t)$ with a real stochastic function $\eta(t)$. The stochastic noise is then constructed from a set of independent random telegraph noise (RTN) sources: 
    $\eta(t)=C\int_{\nu_a}^{\nu_b} \nu_0^{-1/2}\eta_{\mathrm{0}}(t,\nu_0)\dd{\nu_0}$, where $\eta_{\mathrm{0}}(t,\nu_0)$ describes a single RTN that switches between $\pm 1$ with frequency $\nu_0$. Here $C^2=1/\log(\nu_b/\nu_a)$ is a normalizing factor such that $\overline{\eta(t)^2}=1$, and the integrand weight $\nu_0^{-1/2}$ yields a desired noise power spectrum:
    \begin{equation}\label{eq:S_omega_cl}
        S(\omega) = \frac{2\lambda^2C^2}{\omega}\left[
        \arctan\left(\frac{2\nu_b}{\omega}\right)
        -
        \arctan\left(\frac{2\nu_a}{\omega}\right)
        \right]\;.
    \end{equation}
	For $0<\omega\ll \nu_a$, $S(\omega)\propto\mathrm{constant}$; for $\nu_a\ll\omega\ll \nu_b$, $S(\omega)\propto1/\omega$;
	for $\omega\gg\nu_b$, $S(\omega)\propto1/\omega^2$. In our simulation, we simply set $\nu_b = 100\nu_a = \omega_B$, and thus $S(\omega)$ has a soft high-frequency cutoff. It is evident that our filter function cannot effectively filtrate the `long tail' of $S(\omega)$, and no decoupling technique can do so as established by the no-go theorem in Ref.~\cite{Wang_Liu_No_go_PhysRevA.87.042319}.
	Specifically, following the previous argument about the infidelity expansion, the leading terms $\lambda^2 T^{\mu} \omega_B^{\mu-2}$ can have a small $\mu$, due to the integral $\int_{\omega_B}^{\infty} S(\omega)F(\omega,T)\dd{\omega}$ being nonzero in the soft cutoff case. Therefore, the full infidelity does not experience a $\mathcal{O}(T^{2k+2})$ suppression for any $\omega_B T$. 
	Nevertheless, as shown in Fig.~\hyperref[fig:full infid plot]{\ref*{fig:full infid plot}(b)}, our smooth pulse ($k=6$) outperforms the ideal $\mathrm{UDD}_6$ at small $\omega_BT$. This is due to the always-on nature of our pulse, which constantly suppresses the time-dependent perturbation.

	\section{Conclusion}\label{Sec:conclusion}
	We have introduced a protocol to design bounded continuous control fields that produce arbitrary single-axis rotations robust to low-frequency time-dependent noise perpendicular to the control axis. We showed that the leading-order infidelity caused by an arbitrary bath is controlled by the low-frequency behavior of an effective filter function. Our protocol is based on solving a set of integral constraints that enforce the flattening of the filter function at low frequencies. We showed that these integrals can be mapped to a sequence of plane curves such that the vanishing of the integrals translates to closure of the curves. Inspired by this observation, we developed a numerical technique that uses a damped Newton method to obtain robust pulses for a  wide range of target rotations. Moreover, our approach facilitates the design of noise-resistant pulses that respect experimental constraints on the waveform such as non-negativity. We demonstrated the robustness of our smooth pulses by calculating gate fidelities for both classical $1/f$ noise and noise modeled by a random quantum bath. Given the experimental feasibility of the bounded and always-on control pulses introduced in this work, we expect they will be of immediate relevance to the experimental community working on quantum computing and related fields.
	\\
	\section{Acknowledgements}
	
	This work is supported by the U.S. Army Research Office (W911NF-17-0287) and by the U.S. Office of Naval Research (N00014-17-1-2971).

	\appendix
	
	\section{Derivation of Eq.~\eqref{eq:fidelity}}\label{app:derivation_Eq4}
 From Eq.~\eqref{eq:MagnusExpansion}, the leading terms of Magnus expansion are:
 \begin{equation}
 	\left\{
 	\begin{aligned}
 		M_1 &= 
 		\lambda \int_0^T [\sigma_z\cos\phi(t) + \sigma_y\sin\phi(t)]\otimes B_I(t)\mathrm{d}t\\
 		M_1^2&=\lambda^2 \int_0^T\!\!\!\! \mathrm{d}t
 		\int_0^T\!\!\!\! \mathrm{d}s\,
 		e^{-i\sigma_x [\phi(t)-\phi(s)]}\otimes B_I(t) B_I(s)\\
 		M_2 &\propto \lambda^2 \sigma_x
 	\end{aligned}
 \right.
 \end{equation}
To obtain the approximation of Eq.~\eqref{eq:fidelity}, the summation is expanded as:
\begin{widetext}
\begin{equation}\label{eq：trsUsU}
	\begin{aligned}
		&\quad\sum_{\alpha = x,y,z}\!\!
		\mathrm{Tr}[(\sigma_\alpha\otimes \mathbbm{1})
		U_I(T)(\sigma_\alpha\otimes \rho_B)U_I^{\dagger}(T)]\\
		&\approx
		\sum_{\alpha = x,y,z}\!\!
		\mathrm{Tr}[(\sigma_\alpha\otimes \mathbbm{1})
		\left[1- i (M_1+M_2)-\frac{1}{2}M_1^2\right]
		(\sigma_\alpha\otimes \rho_B)
		\left[1+ i (M_1+M_2)^\dagger-\frac{1}{2}M_1^{2\dagger}\right]]\\
		&=
		\sum_{\alpha = x,y,z}\!\!\mathrm{Tr}
		\Big\{
		\mathbbm{1}\otimes \rho_B
		-
		\underbrace{i(M_1+M_2-M_1^\dagger-M_2^\dagger)}_{=0\text{ (traceless)}} \otimes \rho_B
		-
		M_1^2\otimes \rho_B
		+
		(\sigma_\alpha\otimes \mathbbm{1})
		M_1(\sigma_\alpha\otimes \rho_B)
		M_1\Big\} + \mathcal{O}(\lambda^3)\\
		&=6 - 
		3\cdot \underbrace{\mathrm{Tr}[M_1^2\otimes \rho_B]}_{\text{term (1)}}
		+
		\sum_{\alpha = x,y,z}\!\!\underbrace{
			\mathrm{Tr}[(\sigma_\alpha\otimes \mathbbm{1})
			M_1
			(\sigma_\alpha\otimes \rho_B)
			M_1]}_{\text{term (2)}} + \mathcal{O}(\lambda^3)\\
	\end{aligned}
\end{equation}
\end{widetext}
Denote the two-point correlation function as $\langle \hat{B}_I(t)\hat{B}_I(s)\rangle \equiv \mathrm{tr}(\hat{B}_I(t)\hat{B}_I(s)\rho_B)$,
$\text{term (1)}$ is expressed as:
	\begin{equation}\label{eq:term1}
		\begin{aligned}
			&\text{term (1)}=
			2\lambda^2 \int_0^T \!\!\!\! \mathrm{d}t
			\int_0^T\!\!\!\! \mathrm{d}s
			\cos [\phi(t)-\phi(s)]\langle\hat{B}_I(t)\hat{B}_I(s)\rangle
		\end{aligned}
	\end{equation}
As for $\text{term (2)}$, since $M_1$ has the form of $C\sigma_y + D\sigma_z$, then for $\alpha = x$ :
\begin{equation}
	\begin{aligned}
		 \text{term (2)}&= 
		\mathrm{Tr}[(\sigma_x\otimes \mathbbm{1})
		M_1(\sigma_x\otimes \rho_B)M_1]\\
		& = 
		\mathrm{Tr}[(-M_1^2)\otimes \rho_B]
		= (-1)\times\text{term (1)}
		\\
	\end{aligned}
\end{equation}
For $\alpha = y$ and $z$, due to the form of $M_1$, the total contribution is zero:
	$\sum_{\alpha = y,z}\!\!
	\mathrm{Tr}[(\sigma_\alpha\otimes \mathbbm{1})
	M_1
	(\sigma_\alpha\otimes \rho_B)
	M_1]=0$.
Thus, the total fidelity is approximated by:
\begin{equation}
	\begin{aligned}
		\mathcal{F} &\approx \frac{1}{2} + \frac{1}{12}\left[6 - 4\times\text{term (1)}\right]\\
		&=1-\frac{2\lambda^2}{3} \int_0^T \!\!\!\! \mathrm{d}t
		\int_0^T\!\!\!\! \mathrm{d}s
		\cos [\phi(t)-\phi(s)]\langle\hat{B}_I(t)\hat{B}_I(s)\rangle
	\end{aligned}
\end{equation}

\section{Derivation of Eq.~\eqref{eq:S_omg_VQRB}}\label{app:derivation_Eq12}
	Suppose our virtual random quantum bath has $m$ levels ($m\gg 1$), and the coupling operator and the Hamiltonian are separately given by
	$B = W_1/\sqrt{m}$ and $H_B = \omega_B W_2/\sqrt{m}$, where $W_1,W_2$ are random matrices sampled from Gaussian unitary ensemble (GUE).  In this case, the density of eigenvalues $\lambda$ distribution of $W_i/\sqrt{m}$ is given by Wigner's semi-circle law $p(\lambda)=(2\pi)^{-1}\sqrt{4-\lambda^2}$ with $\lambda^2\le 4$, and $p(\lambda)=0$ for otherwise $\lambda$.
	Therefore, assume the quantum bath is in equilibrium with density operator $\rho_B = e^{-\beta H_B}/Z$, the power spectrum density for particular instance of $B$ and $H_B$ is:
	\begin{equation}
		\begin{aligned}
			 &S(\omega) = \frac{\lambda^2}{Z}\int_{-\infty}^{\infty}\!\!\!
			\mathrm{tr}[e^{iH_Bt}Be^{-iH_Bt-\beta H_B}B] e^{-i\omega t }\mathrm{d}t\\
			&=
			\frac{\lambda^2}{Z}\sum_{k,\ell = 1}^m
			\int_{-\infty}^{\infty}\!\!\! |B_{k\ell}|^2 e^{i\omega_B(\lambda_k - \lambda_\ell)t-\beta \omega_B\lambda_\ell}e^{-i\omega t} \mathrm{d}t\\
			&=
			\frac{2\pi \lambda^2}{\omega_B Z}\sum_{k,\ell = 1}^m
			 |B_{k\ell}|^2 e^{-\beta \omega_B\lambda_\ell} \delta\Big(\lambda_k - \lambda_\ell - \frac{\omega}{\omega_B}\Big)\\
		\end{aligned}
	\end{equation}
	In which we obtain the second equal sign via taking the trace under the eigenvector basis of $H_B$, and $B_{k\ell}$ are the matrix entries under this basis.
	Finally, utilizing $\overline{|B_{k\ell}|^2} = \overline{|(W_1)_{k\ell}|^2}/m = 1$ and Wigner's semicircle law, the statistical average of $S(\omega)$ given by: 
	\begin{equation}
		\begin{aligned}
			\overline{S(\omega)}
			&=
			\frac{2\pi \lambda^2}{\omega_B Z}
			\int_{-\infty}^{\infty}\!\!\mathrm{d}\lambda'
			\int_{-\infty}^{\infty}\!\!\mathrm{d}\lambda \,\,p(\lambda')p(\lambda)
			\\&\qquad\times
			e^{-\beta \omega_B\lambda}\delta\Big(\lambda_k - \lambda_\ell - \frac{\omega}{\omega_B}\Big)\\
			&= 
			\frac{2\pi \lambda^2}{\omega_B Z}\int\mathrm{d}\lambda\; p(\lambda)p\Big(\lambda -\frac{\omega}{\omega_B}\Big)e^{-\beta \omega_B \lambda} 
		\end{aligned}
	\end{equation}
where the normalizing factor  is $Z\equiv \int_{-2}^2\dd{\lambda} p(\lambda) e^{-\beta\omega_B \lambda}$ under thermodynamics limit.
The hard cut-off property of $p(\lambda)$ indicates $S(\omega)$ statistically vanishes at $|\omega| \ge 4\omega_B$.
	\\
%

\end{document}